\documentclass[a4paper]{spie}  

\usepackage{amsmath,amsfonts,amssymb}
\usepackage{graphicx}
\usepackage[colorlinks=true, allcolors=blue]{hyperref}
\usepackage[numbers,sort&compress]{natbib}

\usepackage{booktabs}       
\usepackage{subcaption}     

\title{Concept drift of simple forecast models as a diagnostic of low-frequency, regime-dependent atmospheric reorganisation}

\author[a]{Haokun Zhou}
\affil[a]{Imperial College London}

\authorinfo{E-mail: haokun.zhou25@imperial.ac.uk}

\begin{document} 
\maketitle
\begin{abstract}
Data-driven weather prediction models implicitly assume that the statistical relationship between predictors and targets is stationary. Under anthropogenic climate change, this assumption is violated, yet the structure of the resulting concept drift remains poorly understood. Here we introduce concept drift of simple forecast models as a diagnostic of atmospheric reorganisation. Using ERA5 reanalysis, we quantify drift in spatially explicit linear models of daily mean sea-level pressure and 2\,m temperature. Models are trained on the 1950s and 2000s and evaluated on 2020--2024; their performance difference defines a local, interpretable drift metric. By decomposing errors by frequency band, circulation regime and region, and by mapping drift globally, we show that drift is dominated by low-frequency variability and is strongly regime-dependent. Over the North Atlantic--European sector, low-frequency drift peaks in positive NAO despite a stable large-scale NAO pattern, while Western European summer temperature drift is tightly linked to changes in land--atmosphere coupling rather than mean warming alone. In winter, extreme high-pressure frequencies increase mainly in neutral and negative NAO, whereas structural drift is concentrated in positive NAO and Alpine hotspots. Benchmarking against variance-based diagnostics shows that drift aligns much more with changes in temporal persistence than with changes in volatility or extremes. These findings demonstrate that concept drift can serve as a physically meaningful diagnostic of evolving predictability, revealing aspects of atmospheric reorganisation that are invisible to standard deviation and storm-track metrics.
\end{abstract}


\section{Introduction}

Climate change has rendered the atmosphere a fundamentally non-stationary system. Mean states, variability and the spatial structure of teleconnections are all evolving, and these changes inevitably modify the empirical relationships that underpin statistical and machine-learning (ML) prediction models. When such relationships change over time, models trained on past data degrade when applied to new conditions, a phenomenon known in the ML literature as concept drift \citep{Gama2014}. Concept drift has been treated extensively as an algorithmic challenge, but much less as a physical signal that can be diagnosed and interpreted.

Within the ML community, concept drift is typically defined as a change in the joint distribution of inputs and outputs or in the conditional mapping between them, and a large body of work focuses on detection and adaptation methods such as sliding windows, ensemble schemes and explicit drift detectors \citep{Gama2014,Hinder2024}. In Earth-system applications, similar concerns have begun to appear. Environmental and hydrological models have been shown to lose skill as land use, forcing and observation systems change \citep{Rivas2023,Hu2025}, and recent discussions of ML-based weather prediction highlight the vulnerability of purely data-driven models to changes in the climate regime encoded in reanalyses \citep{Schultz2021,Price2023}. In these studies, however, concept drift remains primarily a problem to be mitigated rather than a lens through which to view atmospheric change.

At the same time, the physical climate literature has developed a detailed picture of the dynamical and thermodynamical processes that organise low-frequency variability. Large-scale teleconnections such as the El Ni\~{n}o--Southern Oscillation (ENSO) and the North Atlantic Oscillation (NAO) strongly structure temperature and pressure variability over the Euro-Atlantic and North American sectors, and a substantial body of work has shown that their extratropical impacts are non-stationary on decadal timescales \citep{Greatbatch2004,RodriguezFonseca2016,JimenezEsteve2020,Haszpra2020,Mezzina2020}. In parallel, studies of the eddy-driven jet, storm tracks and blocking have demonstrated that midlatitude extremes are sensitive to subtle changes in jet latitude, storm-track structure and blocking frequency, and that different diagnostics can yield different interpretations of circulation change \citep{Woollings2012,Mbengue2017,Kautz2022}.

Over land, and particularly over Europe, thermodynamic feedbacks further complicate the picture. European summer heatwaves have been shown to be amplified and prolonged by soil-moisture--temperature feedbacks, and regions of strong land--atmosphere coupling have shifted under anthropogenic warming \citep{Seneviratne2006,Fischer2007a,Fischer2007b,Jaeger2011,Whan2015,Vogel2018,Stegehuis2021}. These studies show that the same mean warming can have very different impacts on variability and extremes depending on soil-moisture regimes and the background circulation.

Despite this rich physical context, there remains a conceptual gap between dynamical studies of regimes, teleconnections and land-surface feedbacks on the one hand, and ML-oriented discussions of concept drift on the other. Existing work tells us where and how the atmosphere has changed, but it does not directly tell us where the empirical ``rules'' learned by simple predictive models become obsolete. This paper seeks to close that gap by treating concept drift itself as a physically meaningful diagnostic of non-stationarity in atmospheric dynamics.

Our central idea is that if we train simple, interpretable statistical models on different climate eras and evaluate them on the same recent period, then differences in their performance can be interpreted as evidence that the underlying input--output relationships have changed. By decomposing these performance differences by frequency band, circulation regime and region, we can ask where drift is dominated by slow background shifts, where it is associated with teleconnection reorganisation, and where it is anchored in land-surface feedbacks or circulation regimes. In this way, the concept-drift perspective becomes a bridge between ML robustness questions and physical climate diagnostics.

In what follows, we develop this idea using ERA5 reanalysis, simple spatially aware linear models and a set of standard dynamical and thermodynamical diagnostics. Section~\ref{sec:methods} describes our data and methods. Section~\ref{sec:results} presents four sets of results that together form a coherent picture: low-frequency, regime-dependent drift in midlatitude pressure fields; the relationship between ENSO reorganisation and simple-model drift; the role of land--atmosphere coupling in Western European summer temperature drift; and the decoupling of winter extremes from predictability shifts in the North Atlantic--European sector. In Section~\ref{sec:discussion} we discuss the implications of these findings for both climate interpretation and ML-based weather prediction, before summarising our conclusions in Section~\ref{sec:conclusions}.

\section{Data and methods}
\label{sec:methods}

We base our analysis on the ERA5 reanalysis \citep{Hersbach2020}, focusing on daily mean sea-level pressure (MSLP) and 2\,m air temperature over the Northern Hemisphere extratropics ($20^\circ$--$80^\circ$\,N). To define distinct climate eras, we divide the record into three periods: 1950--1959, 2000--2009 and 2020--2024. The first two periods serve as training decades, while the most recent period serves as a common evaluation window. We restrict our main analyses to regions and seasons where ERA5 is considered reliable back to 1950 and interpret early-period results in light of documented differences in pre- and post-satellite-era reanalysis fidelity.

At each grid point we train two classes of simple daily prediction model. The first class consists of pointwise persistence regressions, which predict next-day MSLP or temperature from the same variable at the same location on the current day. The second class consists of spatial gradient models, which predict central MSLP from a $3 \times 3$ neighbourhood: the central value and its local zonal and meridional gradients. All models are linear and are fitted independently at each grid point by ordinary least squares. For each model class, we estimate separate parameter sets for the 1950s and 2000s using only data from those decades.

We define concept drift by comparing how these models perform on the recent evaluation period. Specifically, for each grid point and model we compute the daily root-mean-square error (RMSE) over 2020--2024, separately for models trained on the 1950s and on the 2000s. The relative difference in RMSE between the ``Past'' and ``Recent'' models, expressed as a percentage of the Recent model’s RMSE, provides a local measure of drift: positive values indicate that a model trained earlier in the record generalises worse to current conditions than a model trained closer in time to the evaluation period. Because both models are evaluated on the same set of days, model error differences can be interpreted as changes in the input--output mapping rather than changes in the forcing.

To understand the temporal structure of drift, we decompose daily MSLP and temperature into three frequency bands using standard filters: a high-frequency band with periods shorter than 2.5 days, a synoptic band between 2.5 and 6 days, and a low-frequency band with periods longer than 6 days. We then compute drift separately for each band by filtering both the predictors and the targets before fitting and evaluating the models. In this way, we can ask whether concept drift arises primarily from changes in transient synoptic systems or from slower background and regime variability.

We further stratify our evaluation by large-scale circulation regimes. For the North Atlantic--European sector we use a standard NAO index to classify each day as negative, neutral or positive NAO, and we compute drift separately for each phase. For global teleconnection analyses we classify winters by ENSO phase using the Oceanic Ni\~{n}o Index and again estimate phase-dependent drift. To connect drift to known teleconnection changes, we compute composite ENSO responses in MSLP and temperature for the 1950s and 2000s and compare their anomalies relative to neutral conditions, including pattern correlations and grid-point significance based on two-sample $t$-tests.

In order to interpret drift in terms of persistence, variability and extremes, we compute for each grid point and decade the lag-1 autocorrelation of MSLP and temperature, the day-to-day variance, and band-pass storm-track variance based on 2.5--6-day filtered MSLP. Over Western Europe we identify extreme winter high-pressure events using a high percentile threshold and track their initiation, duration, spatial extent and peak intensity. We also compute blocking frequency, intensity and location over the North Atlantic--European sector using a conventional MSLP-based blocking index and standard significance tests for changes between decades.

Finally, to investigate the role of land-surface feedbacks in summer temperature drift, we construct a simple proxy for land--atmosphere coupling based on the relationship between soil moisture and temperature anomalies and compare its spatial changes between the 1950s and 2000s. We also identify weak-flow, low-wave-activity regimes over Western Europe by clustering low-wave days in the low-frequency pressure field using $k$-means. Throughout, we use correlation, regression, analysis of variance and effect-size measures (e.g.\ Cohen's $d$, $\eta^2$) to link drift to these diagnostics and apply appropriate statistical tests to assess significance.

\section{Results}
\label{sec:results}

\subsection{Low-frequency, regime-dependent drift in midlatitude pressure fields}

We begin by assessing whether concept drift in simple statistical models is detectable at all and, if so, at what temporal scales. Because the simplest models make the fewest structural assumptions, we first consider pointwise persistence regressions over Western Europe. When we train a persistence model on 1950--1959 and apply it to 2020--2024, we find a small but statistically significant increase in RMSE relative to an otherwise identical model trained on the 2000s: the 1950s-trained model exhibits a 0.09\% higher relative RMSE ($p = 0.026$). This modest degradation already indicates that the local persistence relationship has changed over time, even when spatial context is ignored.

When we move from pointwise persistence to spatial gradient models, the same comparison reveals a clearer signal. For models that use a $3 \times 3$ neighbourhood of MSLP values and gradients, the relative RMSE increase of the 1950s-trained model grows by a factor of 2.78 to 0.25\% ($p = 1.7 \times 10^{-5}$). Thus, once we allow models to exploit spatial patterns, the non-stationarity of those patterns becomes much more apparent in their performance.

This amplification leads naturally to the question of which timescales dominate the observed drift. When we decompose the daily fields into high-frequency, synoptic and low-frequency bands, the answer is unambiguous. In the low-frequency band, relative RMSE differences between 1950s- and 2000s-trained $3 \times 3$ models reach $+3.11\%$ for MSLP and $+3.23\%$ for temperature, compared with $+1.81\%$ in unfiltered data. By contrast, drift in the synoptic band is essentially zero ($-0.02\%$), and high-frequency drift is small and largely confined to summer, with winter mean RMSE differences of $-10^{-6}$\,Pa and summer differences of $1.44 \times 10^{-4}$\,Pa ($p < 0.001$). For these simple models, concept drift is therefore fundamentally a low-frequency phenomenon.

Having established that low-frequency drift is dominant, we next ask how it depends on large-scale circulation regimes. When we stratify daily drift by NAO phase, we find that the degradation of 1950s-trained daily models is significantly greater on positive NAO days than on neutral or negative days ($p = 4 \times 10^{-4}$). This is particularly striking because the large-scale positive-phase NAO pattern itself is highly correlated between the 1950s and 2000s, with a spatial correlation of $r = 0.8852$. The increased drift therefore reflects changes in the local mapping between gradients and central pressure rather than changes in the pattern of gradients.

Indeed, while local gradients show very small shifts between decades ($|\text{Cohen's } d| < 0.08$), the regression coefficients that map those gradients to central pressure change with medium effect sizes. For example, in positive NAO the intercept coefficient shifts with $d = 0.292$, the meridional gradient coefficient weakens by 21.5\% with $d = -0.355$, and the zonal gradient coefficient weakens towards zero by 51.5\% with $d = 0.332$. When we repeat the analysis on low-pass filtered fields, the contrast becomes even sharper: in the $>6$-day band, low-frequency drift is 40.9\% larger in positive NAO than in neutral NAO ($t = 5.18$, $p = 2.5 \times 10^{-7}$) and 42.8\% larger than in negative NAO ($t = 4.97$, $p = 8.8 \times 10^{-7}$), whereas neutral and negative phases differ by only 1.3\% and are statistically indistinguishable ($p = 0.846$).

The regime dependence we observe locally is mirrored at larger scales when we construct a hemispheric map of low-frequency MSLP drift using the same spatial gradient models. On this map, models trained on the 1950s underperform those trained on the 2000s on 65.7\% of Northern Hemisphere extratropical grid points, with a hemispheric mean drift of $+0.0515$\,hPa. The spatial pattern, however, is not uniform. Coherent drift hotspots emerge over the Mediterranean--Middle East, where mean drift lies between 0.13 and 0.17\,hPa and local maxima reach 0.78\,hPa, and secondary hotspots appear over the North Atlantic, North Pacific and western North America with mean drift of around 0.023--0.027\,hPa. By contrast, a notable coldspot appears over the tropical Atlantic and Caribbean, where drift is negative, reaching values as low as $-0.45$\,hPa.

To place these patterns in context, we examine how drift relates to changes in persistence, variance and extremes. Spatially, concept drift correlates moderately with changes in lag-1 autocorrelation of MSLP and temperature, with $r = 0.290$ and 8.4\% of variance explained. By contrast, drift is nearly orthogonal to changes in day-to-day variance, with $r = 0.077$ and only 0.6\% of variance explained. The association with changes in winter extreme high-pressure frequency over Western Europe is weak to moderate ($r = 0.262$, $R^2 = 0.069$), and the association with changes in storm-track variance is moderate and negative ($r = -0.313$, $R^2 = 0.098$), with stronger negative correlations ($r \approx -0.60$) at the storm-track entrance. This comparison indicates that the non-stationarity detected by our simple models is more closely related to changes in the temporal structure of variability than to changes in its amplitude or in the frequency of extremes.

\subsection{Teleconnection reorganisation and simple-model drift}

The dominance of low-frequency drift suggests a connection to large-scale teleconnections, and ENSO is a natural candidate for such a link. Previous studies have shown that ENSO teleconnections to the Euro-Atlantic and North American sectors are themselves non-stationary, with changes in pattern, amplitude and even sign over recent decades \citep{Greatbatch2004,RodriguezFonseca2016,JimenezEsteve2020,Haszpra2020,Mezzina2020,Beverley2024}. Our question is not whether these teleconnections have changed---that is well established---but how much their reorganisation contributes to the concept drift that our simple models exhibit.

To address this question, we first revisit ENSO teleconnection patterns in the ERA5 data. When we construct composite temperature and MSLP anomalies for El Ni\~{n}o and La Ni\~{n}a winters relative to neutral conditions in the 1950s and 2000s, we recover the qualitative picture described in the literature: absolute patterns remain geographically similar, with pattern correlations exceeding 0.94, but anomalies relative to neutral states decorrelate substantially, particularly over Eurasia and North America. In boreal winter, mean pattern correlations of El Ni\~{n}o temperature anomalies relative to neutral conditions between the 1950s and 2000s drop to $r = -0.167$, with local reversals such as up to 9.98\,K colder El Ni\~{n}o anomalies over parts of Kazakhstan and up to 6.21\,K warmer La Ni\~{n}a anomalies over north-western Russia. Mechanistically consistent MSLP changes include strengthened north--south gradients over western Eurasia, with El Ni\~{n}o and La Ni\~{n}a pressure differences of $+2.42$ and $+2.95$\,hPa respectively, and significant North Atlantic La Ni\~{n}a pressure increases of around 930\,Pa. Across variables and phases, between 56\% and 69\% of winter teleconnection-difference grid points exceed conventional significance thresholds ($p < 0.05$).

Over North America, the ENSO--MSLP teleconnection reorganises even more strongly, with winter spatial correlation of El Ni\~{n}o MSLP anomalies between the two decades of $r = -0.531$ and corresponding sign reversals over the south-eastern United States. For La Ni\~{n}a, the pressure pattern correlation is $r = -0.087$, with an $>80\%$ reduction of the western North America high ($-252.9$\,Pa) and a sign reversal over the western North Pacific ($+186.6$\,Pa). These pressure changes translate into a significant winter temperature dipole under El Ni\~{n}o, with cooling of $-2.55^\circ$C in the Pacific Northwest and warming of $+2.25^\circ$C in the south-eastern United States; 84.3\% of grid points show significant changes at $p < 0.05$.

Although these teleconnection changes are physically large, they translate only weakly into hemispheric-scale concept drift for simple 24-hour forecasts. When we compare the drift of 1950s- versus 2000s-trained MSLP models during El Ni\~{n}o and La Ni\~{n}a winters, we find mean relative drift of 0.16\% during El Ni\~{n}o and 0.02\% during La Ni\~{n}a, a difference of only 1.12\,Pa in absolute RMSE, which is not significant in time ($p = 0.205$; Cohen's $d = 0.084$). Spatially, the patterns of drift during El Ni\~{n}o and La Ni\~{n}a differ significantly in a paired sense ($p < 0.001$), but when averaged over the hemisphere they cancel enough that the net temporal signal remains small.

The same is true for temperature. Over North America, the spatial pattern of 24-hour winter temperature drift bears only a weak inverse correlation with the spatial pattern of change in El Ni\~{n}o's winter temperature impact ($r = -0.0798$, $p = 1.29 \times 10^{-35}$, $R^2 = 0.0064$), and changes in ENSO's temperature patterns explain essentially none of the spatial variation in changes in local persistence ($r = -0.0131$, $p = 4.65 \times 10^{-6}$, $R^2 = 0.0002$). Changes in the NAO's influence on volatility are similarly decoupled from persistence change ($r = -0.042$, $R^2 = 0.0017$).

\begin{figure}[htbp]
    \centering
    \begin{subfigure}[b]{0.48\textwidth}
        \centering
        \includegraphics[width=\textwidth]{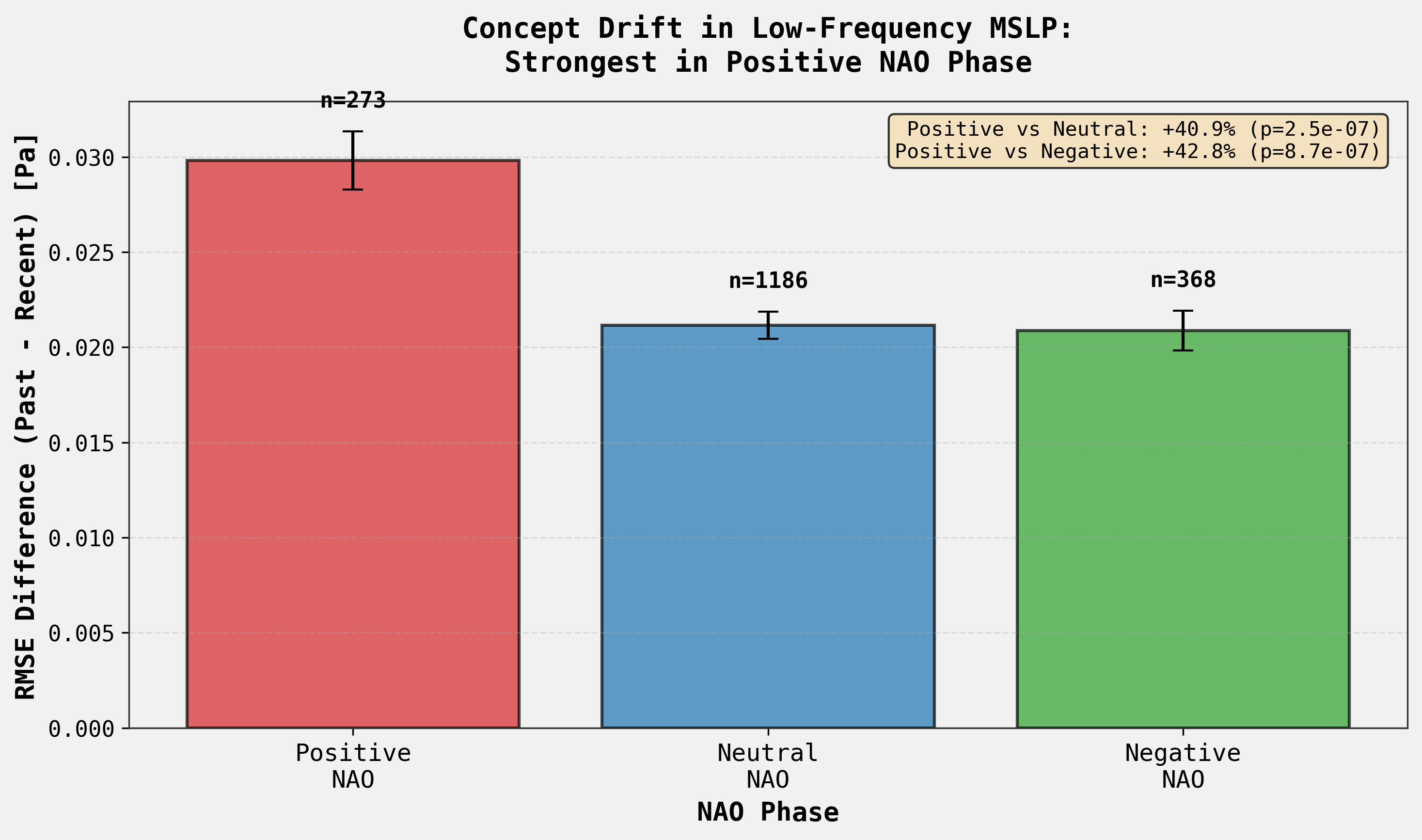}
        \caption{} 
        \label{fig:nao_drift}
    \end{subfigure}
    \hfill
    \begin{subfigure}[b]{0.48\textwidth}
        \centering
        \includegraphics[width=\textwidth]{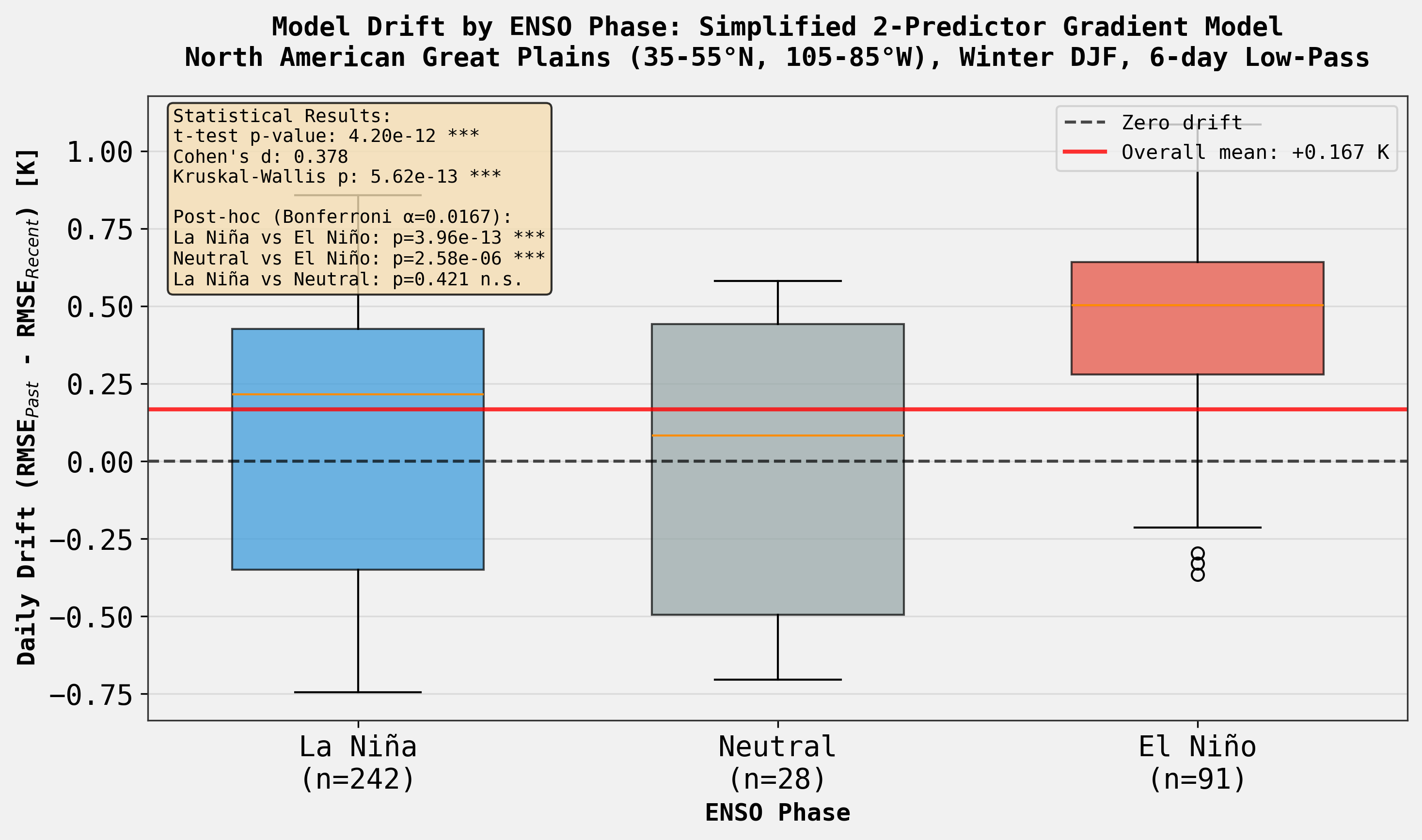}
        \caption{}
        \label{fig:enso_drift}
    \end{subfigure}
    
    \caption{\textbf{Large-scale circulation regimes modulate concept drift.} 
    \textbf{a}, Mean low-frequency MSLP concept drift over the North Atlantic--European sector stratified by NAO phase ($n=1{,}827$ days). Drift is significantly amplified during the positive NAO phase ($+40.9\%$ relative to neutral; $p < 10^{-6}$), indicating that the linear mappings governing zonal flow have degraded more than those governing blocking or meridional flow. Error bars represent the 95\% confidence interval. 
    \textbf{b}, Distribution of winter temperature drift in the North American Great Plains stratified by ENSO phase. Drift is significantly elevated during El Ni\~{n}o winters (mean $+0.167$\,K) compared to La Ni\~{n}a or neutral conditions ($p < 10^{-12}$), reflecting a breakdown in teleconnection stationarity specific to the El Ni\~{n}o base state.}
    \label{fig:regime_modulation}
\end{figure}

These hemispheric results might suggest that ENSO reorganisation is irrelevant for concept drift, but that conclusion would ignore important regional structure. When we focus on specific regions, we do find cases where teleconnections matter. Over the Great Plains, for example, winter temperature drift in a spatial gradient model is strongly modulated by ENSO phase. There, a simplified gradient model yields a significant positive mean winter temperature drift of $+0.167$\,K ($p = 4.20 \times 10^{-12}$), and ENSO phase explains 15.2\% of daily drift variance ($\eta^2 = 0.152$). Mean drift during El Ni\~{n}o winters reaches $+0.457$\,K, compared with $+0.077$\,K during La Ni\~{n}a and $+0.003$\,K during neutral winters. Thus, while ENSO-related reorganisation is not the dominant driver of hemispheric concept drift in our simple models, it can be a primary driver in certain regions.

The contrast between weak hemispheric and strong regional links reinforces a key point: large-scale teleconnection changes can be physically important without automatically dominating the non-stationarity experienced by simple predictive models. For the statistical models we consider, ENSO reorganisation is a necessary part of the physical backdrop, but it is not, on its own, sufficient to explain where and when model rules break down.

\subsection{Summer Western European temperature drift and land--atmosphere coupling}

If ENSO reorganisation is not the main driver of concept drift in our models, then the question naturally arises as to what processes are more important. The strong low-frequency signal and the modest role of variance hint at land-surface feedbacks, particularly over Europe where summer heatwaves have been closely linked to soil-moisture--temperature coupling \citep{Seneviratne2006,Fischer2007a,Fischer2007b,Jaeger2011,Whan2015,Vogel2018,Stegehuis2021}. To investigate this possibility, we turn to Western European summer.

When we restrict our analysis to Western Europe and compare low-frequency drift between seasons, we find that drift is significantly larger in summer than in winter for both MSLP and temperature. For MSLP, a $3 \times 3$ model trained on the 1950s shows a $+2.63\%$ relative RMSE increase in summer compared with $+1.95\%$ in winter ($p = 0.0011$). For temperature, low-frequency drift is 39.6\% greater in summer than in winter ($p < 0.001$), with a Cohen's $d$ of 0.368, indicating a small-to-moderate effect. This seasonal asymmetry indicates that the processes that degrade the stationarity of simple temperature--pressure relationships are more active in summer than in winter, at least over this region.

Within summer, drift is not uniform across days. Instead, daily temperature drift scales with the contemporaneous temperature anomaly: hotter days exhibit larger drift. Across all Western European land grid points, daily low-frequency temperature drift correlates with local temperature anomalies with $r = 0.280$ ($p = 9.33 \times 10^{-10}$). When we divide days into quartiles of temperature anomaly, the hottest quartile exhibits 50.9\% larger mean drift than the coolest quartile ($p = 1.81 \times 10^{-6}$). Non-stationarity is therefore most pronounced precisely on those days when the temperature departures from climatology are greatest.

To connect these patterns to land-surface processes, we examine how the spatial structure of summer low-frequency temperature drift over Western Europe relates to a proxy for land--atmosphere coupling. Over land, the correlation between the spatial pattern of drift and the spatial pattern of change in the coupling proxy between the 1950s and 2000s is $r = 0.626$ ($p = 1.24 \times 10^{-15}$), explaining 39.2\% of variance. The relationship is stronger in southern Europe ($r = 0.326$) than in northern Europe ($r = 0.161$), whereas over nearby oceans the correlation is much weaker ($r = 0.159$, $p \ll 10^{-50}$). Mean summer warming alone explains 34.6\% of drift variance ($r = 0.588$), with a regression slope of $0.391^\circ$C of drift per $1^\circ$C of warming.

After regressing out mean summer warming, the residual drift remains strongly linked to coupling changes over land. For all grid points combined, the correlation between residual drift and coupling change is $r = -0.351$ ($p < 0.001$), while over land alone the relationship strengthens dramatically to $r = -0.905$ ($p = 1.28 \times 10^{-9}$). By contrast, the oceanic correlation is much weaker ($r = -0.350$), and the difference between land and ocean correlations is significant (Fisher’s $z$ test $p = 2.13 \times 10^{-7}$). These spatial relationships point to an interpretation in which changes in soil-moisture--temperature feedbacks are primary drivers of non-stationarity in simple predictive relationships, above and beyond the effects of mean warming.

The role of circulation further clarifies this picture. When we compute a proxy for large-scale wave activity, we find that days with higher wave activity, characterised by more pronounced and organised synoptic systems, tend to show lower drift, with $r = -0.161$ ($p = 5.11 \times 10^{-4}$, $R^2 = 0.026$). Clustering low-wave-activity days in the low-frequency pressure field reveals four weak-gradient, quasi-stationary regimes, each with small pressure ranges of 8--13\,hPa and weak gradients of 0.18--0.22\,hPa per grid point. These regimes co-occur with high-drift days more often than expected by chance (36 observed versus 29 expected days), yet their combined frequency does not change significantly between the 1950s (36.41\%) and 2000s (38.70\%), with $Z = 1.01$ and $\chi^2$ $p = 0.142$. What changes instead are the internal characteristics of these regimes, which, embedded in a warmer, drier land-surface state, now produce different local relationships between pressure and temperature than they did in the past.

Finally, the contrast between MSLP and temperature extremes in summer highlights the different physics at work. Pressure drift is reduced during extreme high and low pressures and largest in near-average pressure states, with an analysis of variance indicating a significant dependence of drift on pressure category ($p = 0.0007$, $\eta^2 = 0.031$). Near-average pressure states also exhibit much longer persistence (lag-1 autocorrelation $r = 0.897$ and mean persistence of 18.6 days) than extreme lows ($r = 0.785$, 5.3 days) or extreme highs ($r = 0.774$, 4.2 days). Temperature drift, by contrast, is amplified during warm extremes, where land--surface feedbacks are strongest. Extremes in one variable can therefore coincide with greater stability in its prediction, while extremes in another coincide with greater instability.

\begin{figure}[htbp]
    \centering
    \begin{subfigure}[b]{0.48\textwidth}
        \centering
        \includegraphics[width=\textwidth]{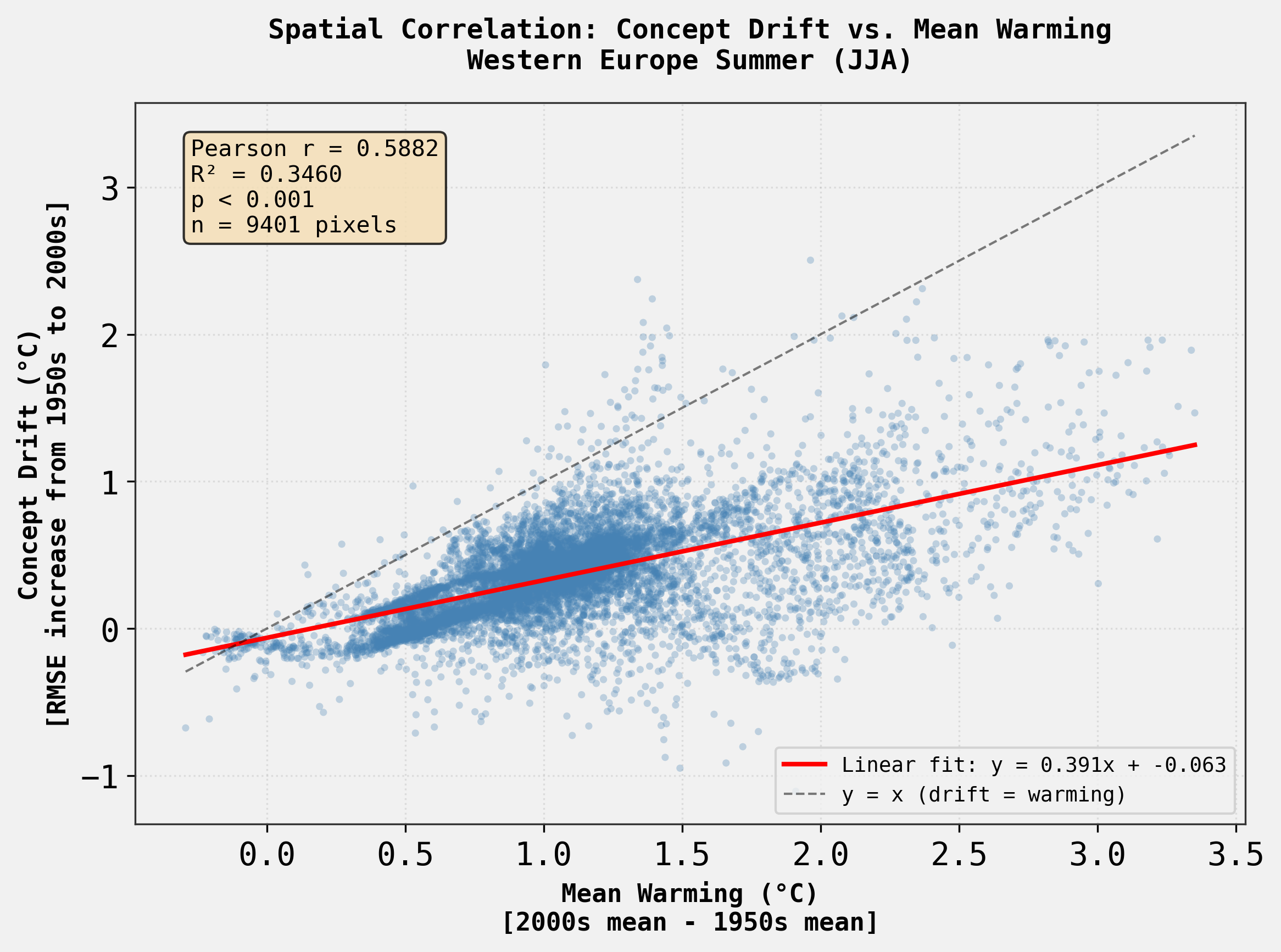}
        \caption{} 
        \label{fig:warming_drift}
    \end{subfigure}
    \hfill
    \begin{subfigure}[b]{0.48\textwidth}
        \centering
        \includegraphics[width=\textwidth]{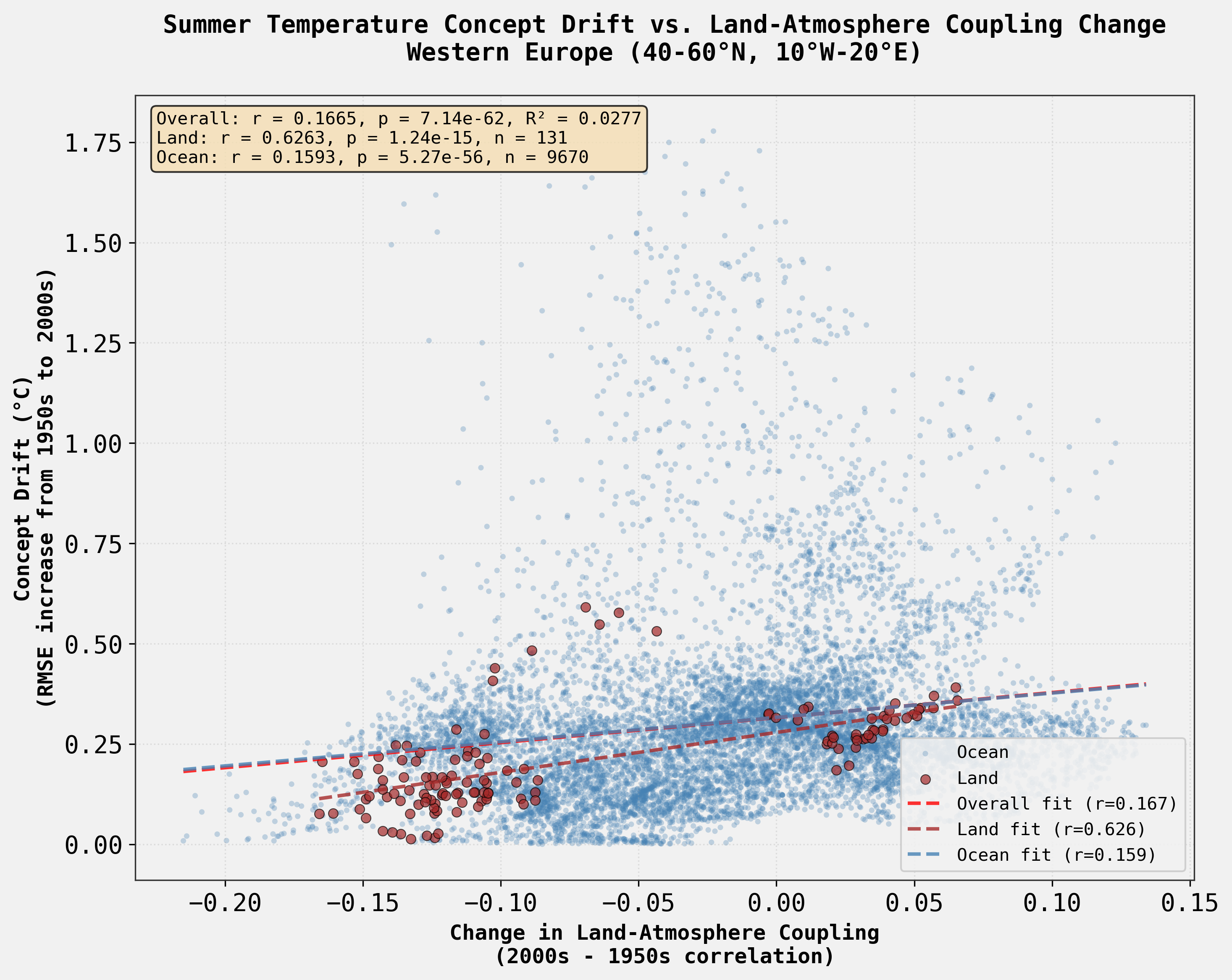}
        \caption{}
        \label{fig:coupling_drift}
    \end{subfigure}
    
    \caption{\textbf{Drivers of concept drift in Western European summer temperatures.} 
    \textbf{a}, Spatial relationship between mean summer warming (2000s minus 1950s) and concept drift (RMSE increase). While drift correlates with warming ($r=0.59$, $p<0.001$), warming magnitude explains only 34.6\% of the spatial variance ($R^2=0.346$), indicating substantial residual non-stationarity. The solid red line indicates the linear fit; the dashed grey line ($y=x$) indicates the theoretical 1:1 scaling. 
    \textbf{b}, Dependence of concept drift on changes in land--atmosphere coupling. Over land grid points (red circles), drift is strongly correlated with coupling changes ($r=0.63$, $p<10^{-15}$), whereas over ocean grid points (blue dots), the relationship is weak ($r=0.16$). This land--ocean contrast demonstrates that thermodynamic feedbacks, rather than atmospheric heating alone, govern the loss of predictability over continental Europe.}
    \label{fig:drivers_drift}
\end{figure}

\subsection{Winter North Atlantic--European regimes: drift versus extremes}

The summer results showed that concept drift and extremes do not have a simple one-to-one relationship. To test more directly whether our drift metric adds information beyond standard extreme-event statistics, we now compare winter low-frequency drift with changes in extreme high-pressure days over the North Atlantic--European sector.

We first quantify how winter extremes have changed. Over Western Europe, the frequency of extreme winter high-pressure days increases by 47\% from the 1950s to the 2000s, despite no significant change in the overall frequency of NAO phases ($\chi^2 = 2.815$, $p = 0.245$). When we stratify by NAO phase, the increase is clearly phase-selective: during neutral NAO, the frequency of extreme high-pressure days rises by 12.18 percentage points (a 47.2\% relative increase, $p < 0.001$), and during negative NAO it rises by 29.72 percentage points (a 93.4\% relative increase, $p = 0.001$). During positive NAO, by contrast, the change is a modest 2.93 percentage points (12.2\% relative) and is not statistically significant ($p = 0.641$). An analyst looking only at extreme-event counts would therefore conclude that neutral and negative NAO have become more ``dangerous'', while positive NAO has changed little.

Low-frequency concept drift paints a different picture. When we compute winter MSLP drift over the same region and stratify by NAO phase, models trained on the 1950s perform worse than models trained on the 2000s on 85.1\% of test days, with a mean RMSE degradation of 2.99\%. This degradation is significant in all phases but is 40.9\% larger in positive NAO than in neutral NAO ($t = 5.180$, $p = 2.53 \times 10^{-7}$) and 42.8\% larger than in negative NAO ($t = 4.967$, $p = 8.75 \times 10^{-7}$), while neutral and negative phases are statistically indistinguishable ($p = 0.846$). In other words, the phase with the largest increase in structural drift is the phase with the smallest change in extreme high-pressure frequency.

Spatial patterns further emphasise this decoupling. The winter drift map reveals a pronounced Alpine hotspot where 1950s-trained models are 4.77 times more likely to exhibit high drift than surrounding regions, and this excess is significant only in winter (Mann--Whitney $p = 0.0008$). Yet this hotspot is not a centre of extreme high-pressure frequency, nor is it preferentially associated with blocking days: mean drift on blocking days is only 9.6\% higher than on non-blocking days, with a Welch's $t$-test $p$-value of 0.3784 and Cohen's $d$ of 0.1094. The local breakdown of linear forecast rules is therefore not simply a mirror of where blocking or extreme anticyclones are most common.

Taken together, these results show that concept drift and extreme-event statistics respond to different aspects of winter circulation change. Extremes increase primarily in neutral and negative NAO through higher initiation rates of anticyclones, whereas low-frequency structural drift is concentrated in positive NAO and in Alpine winter. A diagnostic based solely on extremes would therefore miss some of the regions and regimes where empirical prediction rules have changed most. The drift metric thus provides complementary information, highlighting where model generalisation has degraded even when extreme-event counts appear stable.

\subsection{Benchmarking drift against standard variance diagnostics}

A critical question is whether the concept drift diagnosed here simply captures changes in atmospheric volatility, or whether it identifies a distinct mode of non-stationarity. We therefore benchmark our drift metric against standard variance-based diagnostics.

\begin{figure}[htbp]
    \centering
    \includegraphics[width=\textwidth]{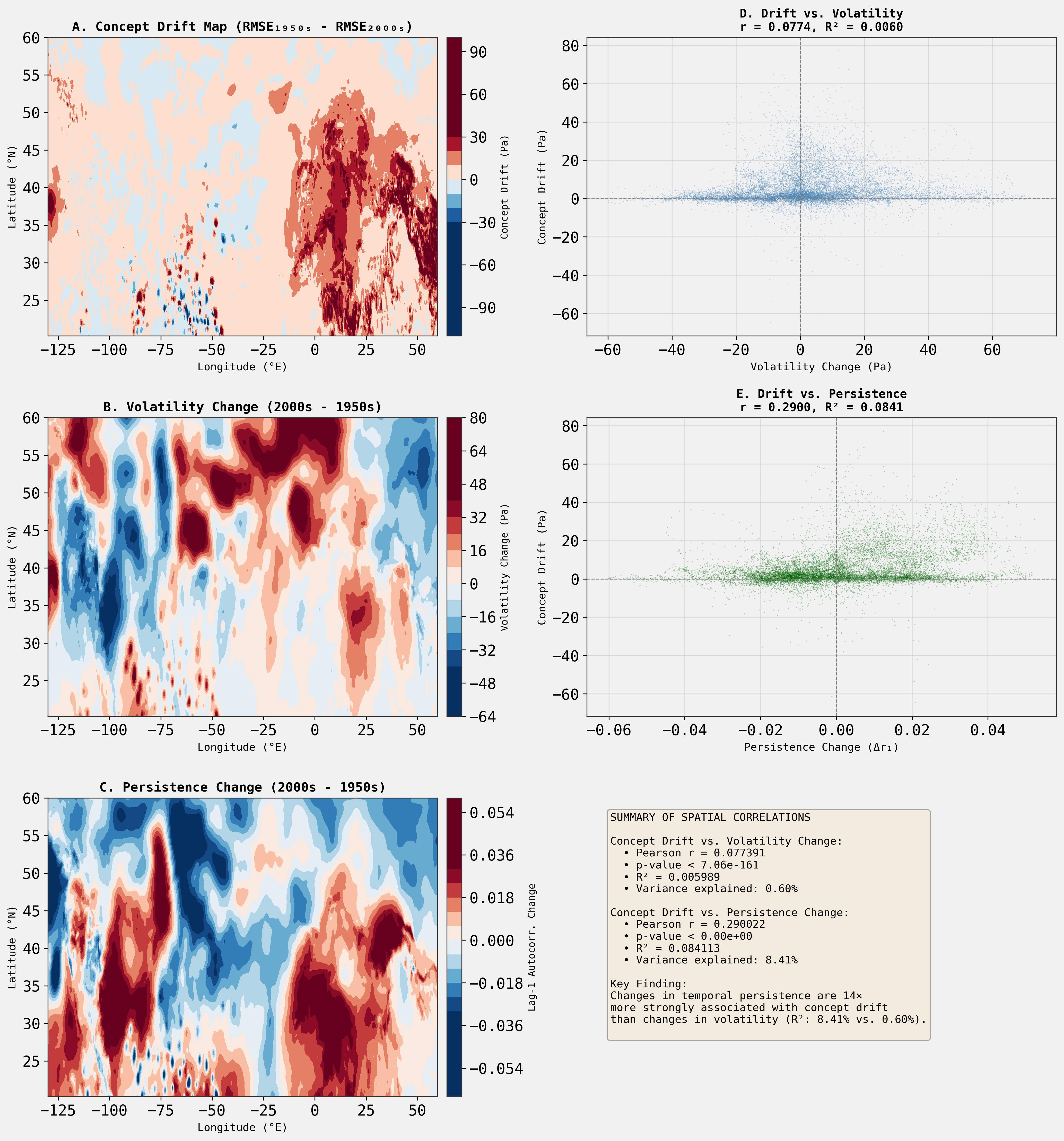}
    \caption{\textbf{Decoupling of concept drift from atmospheric volatility.} \textbf{a--c}, Spatial maps of the change between the 1950s and 2000s in (\textbf{a}) concept drift (RMSE difference), (\textbf{b}) pressure volatility (variance change), and (\textbf{c}) temporal persistence (lag-1 autocorrelation change). Drift hotspots (e.g.\ Mediterranean, North Pacific) do not align with regions of increased volatility. \textbf{d}, Spatial scatter plot of drift vs.\ volatility change showing a negligible relationship ($r = 0.077$, $R^2 < 0.01$). \textbf{e}, Spatial scatter plot of drift vs.\ persistence change showing a moderate positive relationship ($r = 0.29$), indicating that drift is physically linked to changes in system memory rather than system noise.}
    \label{fig:volatility_decoupling}
\end{figure}

We begin with a hemispheric comparison between low-frequency MSLP drift and changes in day-to-day pressure volatility (Fig.~\ref{fig:volatility_decoupling}). We find that low-frequency MSLP drift is effectively orthogonal to changes in volatility. Across the Northern Hemisphere, the spatial correlation between drift and the change in pressure variance is negligible ($r = 0.077$), explaining only 0.6\% of the spatial structure. Consequently, grid points exhibiting the largest drift---and thus the greatest degradation in predictive rules---are rarely those where variance has increased most. For example, the Mediterranean--Middle East drift hotspot and the tropical Atlantic drift coldspot appear in regions where changes in standard deviation are modest. Conversely, changes in storm-track variance show a moderate negative correlation with drift ($r = -0.313$), confirming that regions of intensifying turbulence are not necessarily regions where linear predictive mappings break down.

This decoupling is even more pronounced for Western European summer temperature. While mean warming explains 34.6\% of the spatial variance in drift ($r = 0.588$), changes in day-to-day temperature variability explain almost none of the residual signal. Instead, after regressing out mean warming, residual drift is strongly controlled by changes in land--atmosphere coupling ($r = -0.905$, $p < 10^{-9}$). Standard deviation maps highlight regions of increased thermal amplitude but fail to identify the moisture-limited hotspots where the statistical relationship between today's and tomorrow's state has fundamentally altered.

We quantify this distinction through stepwise regression. For hemispheric MSLP, a model using only variance change predicts drift with an $R^2$ of just 0.006, whereas a model based on persistence change achieves an $R^2$ of 0.084. Adding variance information to the persistence model yields negligible improvement. This demonstrates that concept drift is not a proxy for volatility; rather, it captures structural changes in the temporal mapping---specifically related to persistence and coupling---that standard variance diagnostics miss. Relying solely on changes in standard deviation to assess climate robustness would therefore mask significant areas where empirical prediction rules are becoming obsolete.

\section{Discussion}
\label{sec:discussion}

Our analysis establishes concept drift not merely as an algorithmic challenge, but as a physical diagnostic of atmospheric non-stationarity. By quantifying where simple predictive rules break down, we reveal a pattern of structural decay that is distinct from changes in mean state or variability. This perspective bridges the gap between dynamical climate diagnostics and machine-learning stability, and it yields three main insights into the reorganisation of the midlatitude circulation.

First, the degradation of predictability is regime-dependent rather than uniform. In the North Atlantic, the breakdown of linear mappings is most acute during positive NAO phases, despite the large-scale stability of the NAO pattern itself. The internal ``rules'' governing local pressure gradients are therefore shifting most rapidly in regimes previously considered stable. In summer, the drivers switch from dynamical to thermodynamic: drift over Western Europe is tightly linked to land--atmosphere coupling. The sharp contrast between land (high drift) and ocean (low drift) confirms that soil-moisture feedbacks are actively eroding the stationarity of temperature--pressure relationships, a process that mean warming alone cannot explain.

Second, our benchmarking confirms that concept drift is largely orthogonal to atmospheric volatility. A key result of this study is the lack of spatial correlation between drift and standard variance-based metrics ($R^2 < 1\%$ for hemispheric pressure volatility). Regions experiencing the most rapid loss of predictability---such as the Mediterranean--Middle East hotspot or the Alpine winter sector---are often not the regions experiencing the largest increases in variance. This decoupling is critical: it shows that ``harder to predict'' is not synonymous with ``more variable''. Standard deviation maps, while useful for assessing physical risk, fail to capture the epistemic risk associated with obsolete prediction rules.

Third, the relationship between extremes and predictability is non-linear. Over the Euro--Atlantic sector, we identify a bifurcation in climate risk. Winter extremes (high-pressure anomalies) are increasing primarily in neutral and negative NAO phases, yet structural model degradation peaks in the positive NAO phase. Similarly, in summer, drift is suppressed during pressure extremes but amplified during near-average pressure states, while temperature drift is highest during warm extremes. The atmosphere is therefore not necessarily becoming most unpredictable during its most extreme states; instead, the everyday background physics---the ``normal'' weather---is undergoing a subtle but profound reorganisation that renders historical baselines unreliable.

These insights have direct implications for ML-based weather and climate prediction. The observation that large-scale teleconnection reorganisation does not linearly translate into local concept drift highlights the complexity of downscaling global changes to local scales. For AI weather systems trained on historical reanalyses, our results suggest that training without explicit regime-aware adaptation will leave models particularly vulnerable in identified drift hotspots and regimes. As the climate system moves further from its twentieth-century reference state, drift diagnostics should become standard tools for identifying where the past is no longer a valid prologue to the future.

For climate science more broadly, our study shows that concept-drift maps can complement storm-track metrics, blocking indices and teleconnection analyses. Whereas variance and extremes describe how strongly the atmosphere fluctuates, drift describes where and how the empirical rules of prediction have changed. Incorporating such diagnostics into both model evaluation and climate assessment could improve our understanding of evolving predictability in a warming world.

\section{Conclusions}
\label{sec:conclusions}

In this paper we have used concept drift in simple, spatially aware linear models as a diagnostic of non-stationarity in the midlatitude atmosphere under climate change. By training models on the 1950s and 2000s, evaluating them on 2020--2024, and decomposing their performance differences by frequency band, regime and region, we have shown that drift is dominated by low-frequency variability, structured by circulation regimes and teleconnections, and strongly influenced by land-surface feedbacks.

Over Western Europe and the broader North Atlantic--European sector, we find that low-frequency pressure drift peaks in positive NAO despite a stable large-scale NAO pattern, that summer temperature drift is tightly linked to changes in land--atmosphere coupling rather than to mean warming alone, and that winter extremes and predictability degradation are decoupled across NAO phases. Globally, we find coherent drift hotspots and coldspots that align more closely with changes in persistence than with changes in variance or extremes.

These findings suggest that concept drift can serve as more than an operational nuisance; it can act as a window into how and where the atmosphere's internal ``rules of prediction'' are changing. As ML methods become more deeply integrated into weather and climate services, incorporating such diagnostics into both model design and climate assessment will be essential for understanding and managing the limits of data-driven predictability in a warming world.

\section*{Acknowledgements}

Special thanks to the EdisonScientific platform (Kosmos) for providing key observations.

\end{document}